\documentclass[epj]{svjour}

\usepackage{graphics}
\usepackage[figuresright]{rotating}
\usepackage{amssymb}
\usepackage{bm}
\usepackage{epsfig}
\usepackage{color}
\usepackage{amsmath}
\usepackage{latexsym}

\usepackage{enumerate}
\usepackage{multirow}
\usepackage{array}
\newcolumntype{L}[1]{>{\raggedright\let\newline\\\arraybackslash\hspace{0pt}}m{#1}}
\newcolumntype{C}[1]{>{\centering\let\newline\\\arraybackslash\hspace{0pt}}m{#1}}
\newcolumntype{R}[1]{>{\raggedleft\let\newline\\\arraybackslash\hspace{0pt}}m{#1}}

\newcommand{\beq}{\begin{eqnarray}}
\newcommand{\eeq}{\end{eqnarray}}

\begin{document}

\title{The observational constraints on the flat $\phi$CDM models.}

\author{ {Olga Avsajanishvili}\inst{1} \and {Yiwen Huang}\inst{2} \and {Lado Samushia}\inst{3,1} \and {Tina Kahniashvili}\inst{4,5,1}%
         }
\institute{{Abastumani Astrophysical Observatory, Ilia State University, 3-5 Cholokashvili Ave., Tbilisi, 0194, Georgia} \and%
{Department of Physics, University of California, San Diego, La Jolla, CA 92093, USA} \and
{Department of Physics, Kansas State University, 116 Cardwell Hall, Manhattan, KS, 66506, USA} \and %
{McWilliams Center for Cosmology and Department of Physics, Carnegie Mellon University, 5000 Forbes Ave, Pittsburgh, PA 15213 USA} \and %
{Department of Physics, Laurentian, University, Ramsey Lake Road, Sudbury, ON P3E 2C, Canada}%
}

\authorrunning{Avsajanishvili et al.}
\titlerunning{The observational constraints on the flat $\phi$CDM models}

\date{\today}

\abstract{Most dark energy models have the $\Lambda$CDM as their limit, and if future observations constrain our universe to be close to $\Lambda$CDM Bayesian arguments about the evidence and the fine-tuning will have to be employed to discriminate between the models. Assuming a baseline $\Lambda$CDM model we investigate a number of quintessence and phantom dark energy models, and we study how they would perform when compared to observational data, such as the expansion rate, the angular distance, and the growth rate measurements, from the upcoming Dark Energy Spectroscopic Instrument (DESI) survey. We sample posterior likelihood surfaces of these dark energy models with Monte Carlo Markov Chains while using central values consistent with the Planck $\Lambda$CDM universe and covariance matrices estimated with Fisher information matrix techniques. We find that for this setup the Bayes factor provides a substantial evidence in favor of the $\Lambda$CDM model over most of the alternatives. We also investigated how well the CPL parametrization approximates various scalar field dark energy models, and identified the location for each dark energy model in the CPL parameter space.
}

\maketitle

\section{Introduction}
\label{intro}
It is well established that our universe is undergoing an accelerating
expansion today, Refs. \cite{Riess:1998cb,Perlmutter:1998np,Komatsu:2010fb}.
Several observations suggest that this accelerated expansion started relatively
recently at $z\sim 0.7$, Refs. \cite{Frieman:2008sn,Weinberg:2012es}. One of
the possible explanations is to assume the presence of dark energy as a
dominant component of the total energy density budget in the universe today
(i.e. around 70\% of the  universe matter-energy content today is a substance
with negative pressure that drives today's accelerated expansion). Dark energy
is characterized by an equation of state (EOS) parameter ${w}$ defined by as a
ratio between the pressure ($p$) and the energy density ($\rho$), $w \equiv
p/\rho$. The accelerated expansion requires that $w <-1/3$. Generally speaking
$w$ parameter might be time dependent.  In the framework of the standard cosmological
(concordance) model, dark energy is represented by the cosmological constant
$\Lambda$ (that was originally introduced by Albert Einstein, and it is assumed
to be associated with the vacuum energy density). This cosmological model is
referred as $\Lambda$CDM model, in this case the EOS parameter is constant,
$w=-1$.  The $\Lambda$CDM model is simple and easy to constrain through
observations, but besides good agreements with existing observational data, the
model has a number of shortcomings (the cosmological constant problem, the
coincidence problem, the matter - anti-matter asymmetry, the weakness of
gravity compared to other forces, etc.), Refs.
\cite{Sahni:1999gb,Carroll:2000fy,Sahni:2002kh,Padmanabhan:2002ji,Peebles:2002gy}.  The most
notable of these puzzles is the cosmological constant problem which stems from
the fact that the theoretically expected value (based on quantum field theory
approach and on dimensional arguments) of the cosmological constant associated
energy density is determined by $M_{\rm pl}^4$ (where $M_{\rm pl}=1.2211\cdot10^{19}$~GeV is the
Planck mass), while the actual value (suggested through observational data) is
order of 120 magnitudes lower, Refs.
\cite{Martin:2012bt,Padilla:2015aaa,Copeland:2006wr}. In order to overcome
this (and other) difficulties (the coincidence problem, for example),  {\it
dynamical} dark energy models were proposed, Refs.
\cite{Caldwell:2009ix,Cai:2009zp}, and see Ref. \cite{Bahamonde:2017ize} for a
recent review.

Several large scale structure surveys missions, such as e.g. Dark Energy
Spectroscopic Instrument (DESI), Wide-Field Infrared Survey Telescope (WFIRST)
and Euclid are scheduled to start operating within the next decade. Upon
completion of these missions, very accurate measurements of the expansion
velocity, angular distance and growth rate in the universe to redshifts of
$z\sim 2$  will be obtained, Refs. \cite{Font-Ribera:2013rwa,Levi:2013gra,Aghamousa:2016zmz,Amendola:2012ys,Spergel:2015sza}.  These measurements cumulatively have a very strong
constraining power on the behavior of both dark energy and gravity on large
length scales. If the $\Lambda$CDM model is {\it not} the correct cosmological
model, we should be able to see this in upcoming data. If, however, the
$\Lambda$CDM model or a model very close to it, is the correct model, the
interpretation of the data will be  less straightforward. One reason for this
is that the most viable dark energy models have the $\Lambda$CDM model as their
limit so the Bayesian arguments about the fine-tuning of the extra parameters
will have to be employed. In this work we refer to  a {\it simulated} DESI data and
study how these models would perform when compared to the baseline $\Lambda$CDM
model. The main question we ask is if the $\Lambda$CDM were the correct model of
cosmology would we be able to unambiguously discard alternative models based on
DESI data.

In this paper we investigate a representative family of dark energy models that
are based on the idea of a cosmological scalar field, Refs.
\cite{Starobinsky:1978,Ratra:1987rm,Steinhardt:1999nw,Linder:2007wa,Ludwick:2017tox}. If the
scalar field, $\phi$, has a slowly rolling stage, the energy density associated
with this field can mimic the presence  of the cosmological constant at late
stages.  There are many proposals for the functional form of the
self-interacting potential of the scalar field that are allowed by the current
observational  data, Refs.
\cite{Ferreira:1997hj,Zlatev:1998tr,Brax:1999gp,Sahni:1999qe,Barreiro:1999zs,Albrecht:1999rm,UrenaLopez:2000aj,Caldwell:2005tm,Chang:2016aex,Scherrer:2008be,Dutta:2009dr,Frieman:1995pm,Rakhi:2009qf,Bag:2017vjp}.
In this paper we  consider two types models: the quintessence (dark energy is presented in the form of a canonical scalar fields) and the phantom models (dark energy is presented in the form of a non-canonical scalar field). As of now, there is no consensus on which of these models is
preferable based on the results obtained from the different observations, Refs.
\cite{Suzuki:2011hu,Betoule:2014frx,Ade:2013zuv,Ade:2015xua,Novosyadlyj:2012qu}.
We study the scalar field models with 10 quintessence and 7 phantom potentials
in the Bayesian framework, Refs.
\cite{Heavens:2017hkr,Dhawan:2017leu,Lonappan:2017lzt}. We also limit ourselves
by considering the flat scalar field dark energy (so called $\phi$CDM) models. This is justified by the fact that large
deviations from the spatial flatness of the universe seem to be well
constrained by the CMB data, Ref. \cite{Ade:2015xua}. We have found that under these assumptions a vast
majority of the scalar field dark energy models will be characterized by low
enough Bayes factors to suggest a substantial preference for the $\Lambda$CDM
model.

The paper is organized as follows: in Sec. II we review the dark energy models
(including the scalar field quintessence and phantom models); in Sec. III we
describe observational tests, our results are presented in Sec. IV, and we
conclude in Sec. V.  We use natural units: $c=\hbar=k_B=1$ throughout the
paper.

\section{Dark energy models}
 We will consider two families of scalar field dark energy (flat)
models: the quintessence (canonical) and the phantom scalar field
(non-canonical) models. These models have opposite properties in their
manifestation today: (i) in the range of the EOS parameter values ($w<-1$ for
the phantom field and $-1/3<w<-1$ for the quintessence field); (ii) in the sign
of the kinetic term in the Lagrangian (the negative sign for the phantom field
and the positive one for the quintessence field); (iii) in the dynamics of the
scalar fields (the quintessence field rolls to the minimum its potential, the
phantom field rolls to the "uphill" its potential); (iv)  in the dynamics of
the dark energy density (increases over time for the phantom field and almost
doesn't change over time for the quintessence field); (v) in the forecast for
the future evolution  of the universe: for the phantom models violent future
events (such as big/little/pseudo rips) are predicted, while in the
quintessence models either an eternal expansion or a re-collapse depending on
the spatial curvature of the universe is predicted.

The action associated with the scalar field, $\phi$, is given field by, Refs. \cite{Weinberg:2008zzc}:
\begin{equation}
S=\frac{M_{\rm pl}^2}{16\pi}
\int{d^{4}x\Bigl[\sqrt{-g}
\Bigl(\pm \frac{1}{2}g^{\mu\nu}\partial_\mu\phi\partial_\nu\phi-
V(\phi)\Bigr)\Bigr]},
\label{eq:S}
\end{equation}
where "$+$"  sign before the kinetic term
($g^{\mu\nu}\partial_\mu\phi\partial_\nu\phi/2$) refers to the quintessence
models, while "$-$" stands for the phantom models; $g^{\mu\nu}$ is the
background metric,\footnote{We assume the flat (suggested by current
observations, Ref. \cite{Ade:2015xua}), homogeneous and isotropic universe that
is described by the Friedmann-Lema\^\i tre-Robertson-Walker (FLRW) spacetime
metric, $ds^2 = dt^2 -a^2(t)d{\bf x}^2$, where $a(t)$ is the scale factor
(normalized to be unity today $a_0\equiv a(t_0)$), and $t$ is the physical
time.} and $V(\phi)$ is the self-interacting potential of the scalar field,
$\phi$. The scalar field is assumed to exhibit the negligible spatial
variations, so that the spatial derivatives are small compared to the time
derivatives, and thus we assume the scalar field to be an homogeneous field.

Varying the action Eq.~(\ref{eq:S}), the Klein-Gordon scalar field equation of motion can be obtained, Refs.
\cite{Peebles:1987ek}:
\begin{equation}
\ddot\phi+3\frac{\dot a}{a}\dot\phi \pm \frac{\partial V(\phi)}{\partial \phi}=0,
\label{eq:eqmotion1}
\end{equation}
where again "$+/-$" sign corresponds to the  quintessence/\\phantom model
respectively, the over-dot denotes a derivative with respect to the physical time, $t$.

The energy density and the pressure of the scalar field are expressed, Refs. \cite{Yoo:2012ug}:
\begin{eqnarray}
\rho_\phi = \frac{M_{\rm pl}^2} {32\pi} \Bigl(\pm \dot{\phi}^2/2 + V(\phi) \Bigr),
\label{eq:Rho}  \\
P_\phi  = \frac{M_{\rm pl}^2}{32\pi} \Bigl(\pm \dot{\phi}^2/2 - V(\phi) \Bigr),
\label{eq:P}
 \end{eqnarray}
and the effective EOS parameter for the scalar field is then given by $w_\phi
=  \dfrac{\pm {\dot\phi}^2/2 - V(\phi)}{{\pm \dot\phi}^2/2 + V(\phi)}$. If the
time-derivatives of the scalar field, $\phi$, are small enough to make the
magnitude of the kinetic term small compared to the potential $|\pm
{\dot\phi}^2/2| \ll V(\phi)$ (the ``slow roll'' condition, Ref.
\cite{Starobinsky:1978}), the EOS parameter is very close to negative
one\footnote{More precisely, for the freezing quintessence scalar fields the
EOS parameter is very close to negative one today, while for the thawing
quintessence/phantom scalar fields the EOS parameter  deviates slightly from
minus one in either direction, Ref. \cite{Dutta:2009dr}.} and the  scalar field
behaves like a slowly-time-varying cosmological constant (sometimes the scalar
field is referred as a slowly rolling  scalar field, Ref.
\cite{Peebles:2002gy}). Below we list the potentials of the quintessential and
phantom models considered in this work:

\subsubsection{The quintessence models}
\begin{itemize}
\item[$\bullet$] Ratra-Peebles potential: \\$V(\phi)=V_0M_\mathrm{pl}^2\phi^{-\alpha}$; $\alpha ={\rm const} >0$, Ref. \cite{Ratra:1987rm}
\item[$\bullet$] Ferreira-Joyce potential:\footnote{This potential was investigated earlier by Lucchin and Matarrese, Ref. \cite{LM1985}, as well as by Ratra and Peebles, Ref. \cite{Ratra:1987rm}. Although the complete detailed description of the model was given by Ferreira and Joice, Ref. \cite{Ferreira:1997hj}.}\\ $V(\phi)=V_0\exp(-\lambda\phi/M_{\rm pl})$;  $\lambda = {\rm const}>0$, Ref. \cite{Ferreira:1997hj}

\item[$\bullet$]Zlatev-Wang-Steinhardt potential:\\ $V(\phi)=V_0(\exp({M_{\rm pl}/\phi})-1)$, Ref. \cite{Zlatev:1998tr}

\item[$\bullet$]Sugra potential:\\$V(\phi)=V_0\phi^{-\chi}\exp(\gamma\phi^2/M_{\rm pl}^2)$; $\chi, \gamma={\rm const}>0$, Ref.\\ \cite{Brax:1999gp}

\item[$\bullet$] Sahni-Wang potential:\\$V(\phi)=V_0(\cosh(\varsigma\phi)-1)^g$; $\varsigma={\rm const}>0$,\\ $g={\rm const}<1/2$, Ref. \cite{Sahni:1999qe}

\item[$\bullet$] Barreiro-Copeland-Nunes potential:\\ $V(\phi)=V_0(\exp(\nu\phi) + \exp(\upsilon\phi))$; $\nu$, $\upsilon={\rm const}\geq0$,\\ Ref. \cite{Barreiro:1999zs}

\item[$\bullet$] Albrecht-Skordis potential:\\ $V(\phi)=V_0((\phi-B)^2 + A)\exp(-\mu\phi)$; $A$, $B={\rm const}\geq0$, $\mu={\rm const}>0$, Ref.  \cite{Albrecht:1999rm}

\item[$\bullet$] Ur\~{e}na-L\'{o}pez-Matos potential: \\$V(\phi)=V_0\sinh^m(\xi M_\mathrm{pl}\phi)$; $\xi={\rm const}>0$,\\ $m={\rm const}<0$, Ref. \cite{UrenaLopez:2000aj}

\item[$\bullet$] Inverse exponent potential: \\$V(\phi)=V_0\exp({M_{\rm pl}/\phi})$, Ref. \cite{Caldwell:2005tm}

\item[$\bullet$]Chang-Scherrer potential: \\$V(\phi)=V_0(1+\exp(-\tau\phi))$; $\tau={\rm const}>0$, Ref. \cite{Chang:2016aex}
\end{itemize}

\subsubsection{The phantom models}
\begin{itemize}
\item[$\bullet$] Fifth power potential:\\ $V(\phi)=V_0\phi^5$, Ref. \cite{Scherrer:2008be}

\item[$\bullet$] Inverse square potential: \\$V(\phi)=V_0\phi^{-2}$, Ref. \cite{Scherrer:2008be}

\item[$\bullet$] Exponent potential:\\ $V(\phi)=V_0\exp(\beta\phi)$; $\beta = {\rm const} >0$, Ref. \cite{Scherrer:2008be}

\item[$\bullet$] Quadratic  potential: \\$V(\phi)=V_0\phi^2$, Ref. \cite{Dutta:2009dr}

\item[$\bullet$] Gaussian  potential:\\ $V(\phi)=V_0(1-\exp(\phi^2/\sigma^2))$; $\sigma={\rm const}$, Ref. \cite{Dutta:2009dr}

\item[$\bullet$] Pseudo Nambu-Goldstone boson potential:\\ $V(\phi)=V_0(1-\cos(\phi/\kappa))$; $\kappa = {\rm const} >0$, Ref. \cite{Frieman:1995pm}

\item[$\bullet$] Inverse hyperbolic cosine potential:\\ $V(\phi)=V_0(\cosh(\psi\phi))^{-1}$; $\psi= {\rm const} >0$, Ref. \cite{Rakhi:2009qf}
    \end{itemize}
In both cases of the quintessence and the phantom models, $V_0$  is the model
parameter with the dimension of GeV$^4$. This parameter is obviously related to
the dark energy density parameter today.

\section{Testing dark energy potentials}

\subsection{Model description}
To see how well we will be able to discriminate between these dark energy
scalar field potentials after upcoming dark energy surveys, we generate a set
of the simulated data (theoretical model predictions) for the Hubble expansion
rate, the angular distance, and the growth rate, in the redshift range of
$0.15<z<1.85$ (with $z=1/a - 1$ is the redshift) expected from DESI mission,
Ref. \cite{Font-Ribera:2013rwa}. The measurements are centered around their true values in our fiducial cosmology with the errorbars based on the Fisher matrix predictions. We compute the theoretical expectation for the
angular distance, Hubble parameter, and the growth rate and treat them as
measurements for our mock data set. We use the standard Fisher matrix
predictions for the covariance of these measurements. The real DESI data will,
of course, be a random realization from the likelihood space that doesn't
necessarily sit on top of the maximum likelihood, and the Fisher matrix
predictions tend to overestimate the constraining power of the data. We don't
expect these effects big enough to significantly affect our conclusion.
We fit this synthetic data by using the standard MCMC analysis method to
estimate the multidimensional posterior likelihood of the model parameters.\\
For all dark energy models we compute:

i){\it The Hubble parameter $H(z)$}:

The first Friedmann equation for the flat universe is, Ref. \cite{Peebles:2002gy}:
\begin{equation}
E^2(z) = \Omega_{\rm r,0}(1+z)^{4}+\Omega_{\rm m,0}(1+z)^{3}+
\Omega_{\rm \phi}(z),
\label{eq:Friedmann}
\end{equation}
here $E(z) = H(z)/H_\mathrm{0}$ is the normalized Hubble parameter, and
$H_\mathrm{0}$ is the Hubble parameter today; $\Omega_{i}(z) \equiv
\rho_i(z)/\rho_{\rm cr}$ is the energy density parameter for "i"-th component
(characterized by the energy density, $\rho_i(z)$).\footnote{The critical energy
density today $\rho_{\rm cr} = {3H_0^2(z)}/({8\pi G})$, where $G=M_{\rm
pl}^{-2}$ is the Newton constant. The  current value of the radiation
(relativistic component) and the matter (non-relativistic component) density
parameter are designated as $\Omega_{\rm r,0}$ and $\Omega_{\rm m,0}$
respectively; $\Omega_{{\rm \phi}}(z)$ is the dark energy density time
dependent parameter. We denote $\Omega_{\rm \phi, 0}\equiv\Omega_{{\rm
\phi}}(z=0)$. The condition of the background metric flatness is given,
$\Omega_{\rm r,0} + \Omega_{\rm m,0} = 1- \Omega_{\rm \phi, 0}$.  }

ii){\it The angular diameter distance}

Assuming a flat universe, the angular diameter distance is given by, Ref. \cite{Weinberg:2008zzc}:
\begin{equation}
d_A(z)=\frac{1}{H_\mathrm{0}(1+z)}\int_0^z\frac{dz'}{E(z')}
\end{equation}

iii) {\it The combination of the growth rate and the matter power spectrum amplitude,
$f(a)\sigma_8(a)$}

The growth rate is given as,
$f(a)=\mathrm{d}\mathrm{ln}D(a)/\mathrm{d}\mathrm{ln}a$, where $D(a)$ is the
growth function defined through the ratio of overdensities, $\delta(a)$, at
different scale factors, as $D(a)=\delta(a)/\delta(a_0)$, normalized to be
unity today, ($D(a_0)=1$), and it is a solution of the following linear
perturbation equation, Ref. \cite{Pace:2010sn}:
\begin{equation}
D^{''}+\Bigl(\frac{3}{a}+\frac{E^{'}}{E}\Bigr)D^{'}-\frac{3\Omega_\mathrm{m,0}}{2a^{5}E^{2}}D=0,
\label{eq:deltaeq}
\end{equation}
here a prime denotes a derivative with respect to the scale factor, $a$,
($^\prime = d/da$).  The matter power spectrum amplitude can be characterized
through the  $\sigma_8(a)$ function,  $\sigma_8(a)\equiv D(a)\sigma_8$, where
$\sigma_8 \equiv \sigma_8(a_0)$  is the rms linear fluctuation  in the mass
distribution on scales $8h^{-1}$ Mpc (with $h$ is the today Hubble constant in
units of 100~km$/$s/Mpc) today.  We fix the value of $\sigma_8$ to its current
best-fit $\Lambda CDM$ value of $\sigma_8=0.815$  from the Plank 2015 data,
Ref. \cite{Ade:2015xua}, (see Ref. \cite{LHuillier:2017ani} for
model-independent cosmological constraints on $\sigma_8$ from growth and
expansion).

The EOS parameter of the dark energy models is often characterized by the
Chevallier-Polarsky-Linder (CPL)\\$w_0-w_a$ parametrization, Ref.
\cite{Chevallier:2000qy,Linder:2002et}:
\begin{equation}
 w(a) = w_0 + w_a(1-a),
\label{eq:EOS}
\end{equation}
where $w_0=w(a=1)$ and $w_a=-a^{-2}({\rm d} w/{\rm d}a)|_{a=1/2}$.

This parametrization fits the EOS parameters for most of the dark energy models
well enough for some effective values of $w_0$ and $w_a$, but may fail to
describe the arbitrary dark energy models to a good precision (few percents)
over a wide redshift range.\footnote{Dark energy is sometimes characterized by
the EOS parameter only, and the corresponding cosmological model is referred as
$w$CDM model, Ref. \cite{Barger:2006vc}},

In addition, the structure growth (in the most dark energy models) tends to be
sensitive (only) to the fractional matter density, $\Omega_{\rm m}(a) =
\Omega_{\rm m}/E^2(a)$, with $\Omega_{\rm m}=\Omega_{\rm m,0}a^{-3}$
and as a consequence, the matter perturbation growth rate function, $f(a)$, with
high accuracy can be parameterized as, Ref. \cite{Wang:1998gt}:
\begin{equation}
 f(a) \approx [\Omega_{\rm m}(a)]^{\gamma(a)},
 \label{eq:f1f2}
\end{equation}
where $\gamma(a)$ is so called {\it the growth index}, and in general it is a
time-dependent function. \footnote{For the $\Lambda$CDM model the effective values of the growth index, $\gamma(a)$, have a quasi-constant behavior, Ref. \cite{Polarski:2016ieb}. For the scalar field models the growth index, $\gamma(a)$,  changes in the narrow interval of its values, Ref. \cite{Polarski:2016ieb,Avsajanishvili:2015elt}.} In the case of the $w$CDM models (or any dark energy
models which are the well approximated by the $w_0-w_a$ parametrization), the
growth index, $\gamma(a)$, scale factor dependence on can be determined from
Eq.~(\ref{eq:f1f2}), see Ref.~\cite{Wu:2009zy}:
\begin{equation}
\gamma(a) = \frac{\ln f(a)}{\ln \Omega_{\rm m}(a)}.
\label{eq:gamma_fun}
\end{equation}
On the other hand, the function, $\gamma(a)$, can be parameterized by a scale
factor independent manner, so called the Linder $\gamma$-parametrization, see
Ref.~\cite{Linder:2007hg}:
\begin{eqnarray}
\gamma=\left\{\begin{array}{rl}
    0.55+0.05(1+w_0+0.5w_a),& \mbox {if } w_0 \ge-1;\\
    0.55+0.02(1+w_0+0.5w_a),& \mbox {if } w_0<-1.
        \end{array}\right.\
\label{eq:Lgamma}
\end{eqnarray}
This parametrization is accurate up to redshift of $z=5$ ($a=0.2$), Ref.
\cite{Avsajanishvili:2015elt}. The numerical value of the $\gamma$ itself
depends on the dark energy model characteristics ($w$-parameter), being equal
to 0.55 for the $\Lambda$CDM model, Ref. \cite{Linder:2007hg}.

We don't use the Linder $\gamma$-parametrization or CPL one in our MCMC chains. Instead we fit
directly to the model predictions by solving the fundamental differential
equations. We do however, as an independent exercise, check how well these
parametrizations work for the dark energy models that we consider. We find that all dark energy models under consideration can be approximated very well by
these two parametrizations.

\subsection{The definition of the starting points for the MCMC chains}
To find the starting points for our MCMC chains, we solve jointly the scalar
field equation for the quintessence and phantom models, Eq.~(\ref{eq:eqmotion1}), the Friedmann equation, Eq.~(\ref{eq:Friedmann}), and
the linear perturbation equation, Eq.~(\ref{eq:deltaeq}), for a wide range of the
free parameters and the initial conditions for matter dominated epoch.  For
each potential we have found the plausible solutions, for which the following
three criteria were simultaneously fulfilled:
\begin{enumerate}
\item
The transition between the matter and dark energy equality
($\Omega_\mathrm{m}=\Omega_\mathrm{\phi}$) happens relatively recently $z\in
(0.6 - 0.8)$, Ref. \cite{Arkhipova:2014dha}.

\item
The matter perturbation growth rate, $f(a)$, and the fractional matter density,
$\Omega_{\rm m}(a)$, are parameterized  by the Linder $\gamma$-parametrization (Eq.~(\ref{eq:Lgamma})).

\item
 The EOS parameter predicted by the different dark energy models should be in
 the agreement with the expected EOS parameter value today (for the phantom
 models $w_0<-1$; for the quintessence  models with $-1<w_0<-0.75$: for the
 freezing type $w_a<0$ and for the thawing type $w_a>0$).
\end{enumerate}

For all  potentials we found the range for (i) the allowed initial conditions
and (ii) the model parameters, which we then used as the starting points for
the MCMC chains.

This is done to make sure that the MCMC chains converge faster by starting them close to the peak of the posterior likelihood. The actual likelihood surface from the converged MCMC chains of course doesn't depend on the starting point.

\section{Results}
We computed the projected covariance matrix of $D_A(z)$, $H(z)$, and $f\sigma_8(z)$ measurements following standard Fisher matrix approach described in Ref. \cite{Font-Ribera:2013rwa}. We assumed 14000 sq. deg. of sky coverage and wavenumbers up to $k_\mathrm{max} = 0.2\ \mathrm{Mpc}/h$. Our variances matched the numbers in Table V of Ref. \cite{Font-Ribera:2013rwa}. We also accounted for covariances between the measurements within the same redshift bin. $D_A(z)$ and $H(z)$ measurements are negatively correlated by approximately 40\%, while correlations with $f\sigma_8(z)$ are below 10\% for all redshift bins.

All dark energy models considered in this work have the following free
parameters,  $\Omega_\mathrm{m,0}$ and $H_\mathrm{0}$. In addition, the scalar
field models have the extra parameters describing the strength and shape of the
potential, $V(\phi)$. These free parameters along with the prior ranges
considered in our MCMC runs are presented in the Tables \ref{table:QP} and
\ref{table:PP}. We have found these  priors using the phenomenological method,
which is described in the previous section, i.e. they correspond to the three
conditions imposed on the solutions for each potential. We have explicitly
checked that most of the high likelihood regions are inside these priors in a
way that the parameter constraints will not be effected by adjusting the prior
ranges.

\begin{table*}[h!]
\begin{tabular}{|m{7cm}|m{3cm} m{3cm}|}
\hline 
\multicolumn{1}{|C{7cm}|}{The quintessence potentials} & \multicolumn{2}{C{8cm}|}{Free parameters} \\
\hline
\hline
 $V(\phi)=V_0M_\mathrm{pl}^2\phi^{-\alpha}$  &   $H_\mathrm{0}(50\div90)$\par
                                                 $\Omega_\mathrm{m0}(0.25\div0.32)$ &
                                                 $V_0(3\div5)$ \par
                                                 $\alpha(10^{-6}\div0.7)$ \\

\hline
$V(\phi)=V_0\exp(-\lambda\phi/M_{\rm pl})$ & $H_\mathrm{0}(50\div90)$\par $\Omega_\mathrm{m0}(0.25\div0.32)$ \par $V_0(10\div10^3)$ & $\lambda(10^{-7}\div10^{-3})$ \par  $\phi_0(0.2\div1.6)$ \par $\dot{\phi_0}(79.8\div338.9)$\\

\hline
$V(\phi)=V_0(\exp({M_{\rm pl}/\phi})-1)$  &   $H_\mathrm{0}(50\div90)$\par $\Omega_\mathrm{m0}(0.25\div0.32)$ \par  $V_0(10\div10^2)$ & $\phi_0(1.5\div10)$ \par $\dot{\phi_0}(350\div850)$\\

\hline
$V(\phi)=V_0\phi^{-\chi}\exp(\gamma\phi^2/M_{\rm pl}^2)$   & $H_\mathrm{0}(50\div90)$\par $\Omega_\mathrm{m0}(0.25\div0.32)$ \par $V_0(10^{-2}\div10^{-1})$ \par $\chi(4\div8)$ & $\gamma(6.5\div7)$ \par  $\phi_0(5.78\div10.55)$ \par $\dot{\phi_0}(680.6\div879)$\\

\hline
$V(\phi)=V_0(\cosh(\varsigma\phi)-1)^g$ &  $H_\mathrm{0}(50\div90)$\par $\Omega_\mathrm{m0}(0.25\div0.32)$ \par   $V_0(5\div8)$ \par  $\varsigma(0.15\div1)$ &  $g(0.1\div0.49)$ \par  $\phi_0(1.8\div5.8)$ \par $\dot{\phi_0}(360\div685)$\\

\hline
$V(\phi)=V_0(\exp(\nu\phi) + \exp(\upsilon\phi))$ & $H_\mathrm{0}(50\div90)$\par $\Omega_\mathrm{m0}(0.25\div0.32)$ \par $V_0(1\div12)$ & $\nu(6\div12)$ \par  $\phi_0(0.014\div1.4)$ \par $\dot{\phi_0}(9.4\div311)$\\

\hline
$V(\phi)=V_0((\phi-B)^2 + A)\exp(-\mu\phi)$  & $H_\mathrm{0}(50\div90)$\par $\Omega_\mathrm{m0}(0.25\div0.32)$ \par $V_0(40\div70)$ \par $A(1\div40)$ & $B(1\div60)$ \par $\mu(0.2\div0.9)$ \par $\phi_0(5.8\div8.45)$ \par  $\dot{\phi_0}(681\div804.5)$\\

\hline
$V(\phi)=V_0\sinh^m(\xi M_\mathrm{pl}\phi)$ & $H_\mathrm{0}(50\div90)$\par $\Omega_\mathrm{m0}(0.25\div0.32)$ \par $V_0(1\div10)$ \par $m(-0.1\div-0.3)$  & $\xi(10^{-2}\div1)$ \par $ \phi_0(0.5\div2.5)$ \par $\dot{\phi_0}(190\div367)$\\

\hline
$V(\phi)=V_0\exp({M_{\rm pl}/\phi})$ & $H_\mathrm{0}(50\div90)$\par $\Omega_\mathrm{m0}(0.25\div0.32)$ \par $V_0(10^2\div10^3)$ & $ \phi_0(5.78\div10.55)$ \par $\dot{\phi_0}(680.6\div879)$\\

\hline
$V(\phi)=V_0(1+\exp(-\tau\phi))$ & $H_\mathrm{0}(50\div90)$\par $\Omega_\mathrm{m0}(0.25\div0.32)$ \par $V_0(1\div10^2)$ & $\tau(10\div10^2)$ \par $\phi_0(0.01\div0.075)$ \par $\dot{\phi_0}(9.4\div32)$\\

\hline
\end{tabular}
\caption{\rm The list of the dark energy quintessence potentials and the free parameters.}
\label{table:QP}
\end{table*}

\bigskip

\begin{table*}[h!]
\begin{tabular}{|m{7cm}|m{3cm} m{3cm}|}
\hline 
\multicolumn{1}{|C{7cm}|}{The phantom potentials} & \multicolumn{2}{C{8cm}|}{Free parameters} \\
\hline

\hline
$V(\phi)=V_0\phi^5$ &  $H_\mathrm{0}(50\div90)$\par $\Omega_\mathrm{m0}(0.25\div0.32)$ \par $V_0(10^{-3}\div10^{-2})$ & $\phi_0(3.37\div3.94)$ \par $\dot{\phi_0}(523\div563.6)$\\

\hline
 $V(\phi)=V_0\phi^{-2}$ & $H_\mathrm{0}(50\div90)$\par $\Omega_\mathrm{m0}(0.25\div0.32)$ \par $V_0(30\div50)$ & $\phi_0(2.83\div5.15)$ \par $\dot{\phi_0}(471.4\div600)$\\

\hline
$V(\phi)=V_0\exp(\beta\phi)$ &   $H_\mathrm{0}(50\div90)$\par $\Omega_\mathrm{m0}(0.25\div0.32)$ \par $V_0(1\div20)$ & $\beta(0.08\div0.3)$ \par $ \phi_0(0.2\div9.14)$ \par $\dot{\phi_0}(79.8\div830.9)$\\

\hline
$V(\phi)=V_0\phi^2$ &  $H_\mathrm{0}(50\div90)$\par $ \Omega_\mathrm{m0}(0.25\div0.32)$ \par $V_0(1\div20)$ & $\phi_0(0.67\div2.8)$ \par $\dot{\phi_0}(191\div450)$\\

\hline
$V(\phi)=V_0(1-\exp(\phi^2/\sigma^2))$ &   $H_\mathrm{0}(50\div90)$\par $ \Omega_\mathrm{m0}(0.25\div0.32)$ \par $V_0(5\div30)$ & $\sigma(5\div30)$ \par $\phi_0(0.67\div2.8)$ \par $\dot{\phi_0}(191\div450)$\\

\hline
$V(\phi)=V_0(1-\cos(\phi/\kappa))$ &  $H_\mathrm{0}(50\div90)$\par $ \Omega_\mathrm{m0} (0.25\div0.32)$ \par $V_0(1\div4)$ & $\kappa(1.1\div2)$ \par $ \phi_0(2.3\div3.37)$ \par $ \dot{\phi_0}(420\div500)$\\

\hline
 $V(\phi)=V_0(\cosh(\psi \phi))^{-1}$ & $H_\mathrm{0}(50\div90)$\par $ \Omega_\mathrm{m0}(0.25\div0.32)$ \par $V_0(10^{-3}\div10^2)$ & $ \psi(10^{-3}\div1)$ \par $  \phi_0(1.4\div2.3)$ \par $\dot{\phi_0}(310\div420.7)$\\

\hline
\end{tabular}
\caption{\rm The list of the dark energy phantom potentials and the free parameters.}
\label{table:PP}
\end{table*}

\section{The Bayesian statistics}

The reconstruction (and constraining) of the dark energy potentials with
minimal priors is a challenging task (see e.g. Ref. \cite{Sangwan:2017kxi} for
more details). To assess the quality of the different models and to
distinguish them from each other, we have applied the Akaike information
criterion ($AIC$), Ref. \cite{Akaike:1974} and the Bayesian (or Schwarz)
information criterion ($BIC$), Ref. \cite{Schwarz:1978}. The information
obtained by these criteria complement each other.

$AIC$ and $BIC$ are defined respectively as,
\begin{equation}
AIC = -2\ln{\mathcal{L}_{max}} + 2k,
\end{equation}
and
\begin{equation}
BIC = -2\ln{\mathcal{L}_{max}} + k\ln{N},
\end{equation}
where $\mathcal{L}_{max}\propto \mathrm{exp}(-\chi^2_\mathrm{min}/2)$ is a
maximum value of the likelihood function; $N$ is a number of free parameters;
$k$ is a number of data points.

We also computed the evidence integral defined as,
\begin{equation}
\mathcal{E} = \int \mathrm{d}^3\bm{p} \mathcal{P}(\bm{p}),
\end{equation}
where $\bm{p}$ are all parameters of the model, $\mathcal{P}$ is the posterior
likelihood (proportional to the local density of MCMC points), and the
boundaries of the integral are given by the prior. We explored how tight the
prior on the extra parameters needs to be for them to be competitive (in the
sense of the Bayesian evidence) with the standard $\Lambda$CDM model.

We explicitly checked that the priors incorporate most of the high posterior
area. Since all dark energy models have the $\Lambda$CDM model as their limit,
ruling them out simply based on the posterior is technically speaking
impossible. Since the synthetic data was generated in the $\Lambda$CDM model
that limit will always result in high likelihood, and because of the finite
size of the errorbars there will always be a region around the best-fit
$\Lambda$CDM model that is consistent with the data. One could however appeal
to the Bayesian evidence and argue that the extra parameters need to be
extremely fine tuned. We numerically integrated the posterior likelihood to get
for all models.

These results are presented in the Tables \ref{table:bicq} and
\ref{table:bicp}.  All these numbers are normalized with respect to the
fiducial $\Lambda$CDM model.
\begin{table*}[t]
\begin{center}
\begin{tabular}{|l|l|l|l|}
\hline
\hspace{0.5 cm}\textrm{Quintessence potentials} & \textrm{AIC}  & \textrm{BIC} & \textrm{Bayes factor}\\
\hline\hline
\hline
 $V(\phi)=V_0M_\mathrm{pl}^2\phi^{-\alpha}$&10&18.7&0.5293\\
[0.2cm]
\hline
$V(\phi)=V_0\exp(-\lambda\phi/M_{\rm pl})$&12&22.4&0.0059\\
[0.2cm]
\hline
$V(\phi)=V_0(\exp({M_{\rm pl}/\phi})-1)$&10&18.7&0.0067\\
[0.2cm]
\hline
$V(\phi)=V_0\phi^{-\chi}\exp(\gamma\phi^2/M_{\rm pl}^2)$&14&26.2&0.0016\\
[0.2cm]
\hline
$V(\phi)=V_0(\cosh(\varsigma\phi)-1)^g$&14&26.2&0.0012\\
[0.2cm]
\hline
$V(\phi)=V_0(\exp(\nu\phi) + \exp(\upsilon\phi))$&14&26.2&0.0053\\
[0.2cm]
\hline
$V(\phi)=V_0((\phi-B)^2 + A)\exp(-\mu\phi)$&16&29.9&0.0034\\
[0.2cm]
\hline
$V(\phi)=V_0\sinh^m(\xi M_\mathrm{pl}\phi)$&14&26.2&0.0014\\
[0.2cm]
\hline
$V(\phi)=V_0\exp({M_{\rm pl}/\phi})$&10&18.7&0.0077\\
[0.2cm]
\hline
$V(\phi)=V_0(1+\exp(-\tau\phi))$&12&22.4&0.0024\\
[0.2cm]
\hline
\end{tabular}
\caption{\rm The list of the dark energy quintessence potentials, with corresponding $AIC$, $BIC$, and Bayes factors.}
\label{table:bicq}
\end{center}
\end{table*}

\begin{table*}[t]
\begin{center}
\begin{tabular}{|l|l|l|l|}
\hline 
\hspace{0.2 cm}\textrm{Phantom potentials} &\textrm{AIC}  &\textrm{BIC} & \textrm{Bayes factor}\\
\hline\hline
\hline
$V(\phi)=V_0\phi^5$ &10 &18.7 &0.0921\\
[0.2cm]
\hline
$V(\phi)=V_0\phi^{-2}$  &10 &18.7 &0.0142\\
[0.2cm]
\hline
$V(\phi)=V_0\exp(\beta\phi)$ &22.4 &12  &0.0024\\
[0.2cm]
\hline
$V(\phi)=V_0\phi^2$  &10& 18.7& 0.0808\\
[0.2cm]
\hline
$V(\phi)=V_0(1-\exp(\phi^2/\sigma^2))$  &12& 22.4& 0.0113\\
[0.2cm]
\hline
$V(\phi)=V_0(1-\cos(\phi/\kappa))$  &12 &22.4 &0.0061\\
[0.2cm]
\hline
 $V(\phi)=V_0(\cosh(\psi \phi))^{-1}$  &12 &22.4 &0.0056\\
[0.2cm]
\hline
\end{tabular}
\caption{\rm The list of the dark energy phantom potentials, with corresponding $AIC$, $BIC$, and Bayes factor values.}
\label{table:bicp}
\end{center}
\end{table*}

\subsection{The $\phi$CDM potentials vs CPL parametrization}
\begin{figure}[t]
\resizebox{0.5\textwidth}{!}{
 \includegraphics{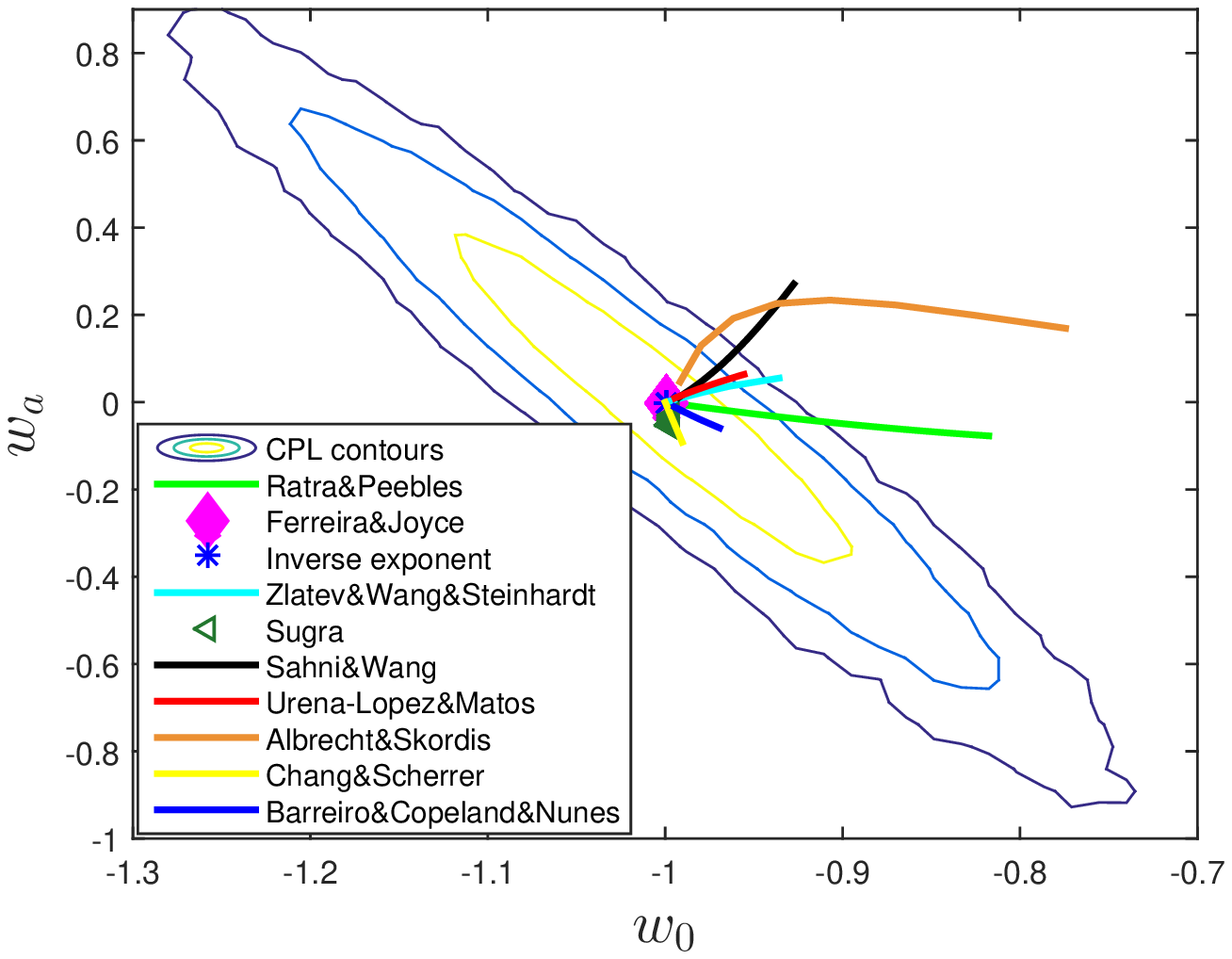}}
 \caption{The comparison of the possible $(w_0,w_a)$ values of the quintessence dark energy potentials with the CPL-$\Lambda$CDM 3$\sigma$ confidence level contours.}
\label{fig:f1}
\end{figure}

\begin{figure}
\resizebox{0.5\textwidth}{!}{
\includegraphics{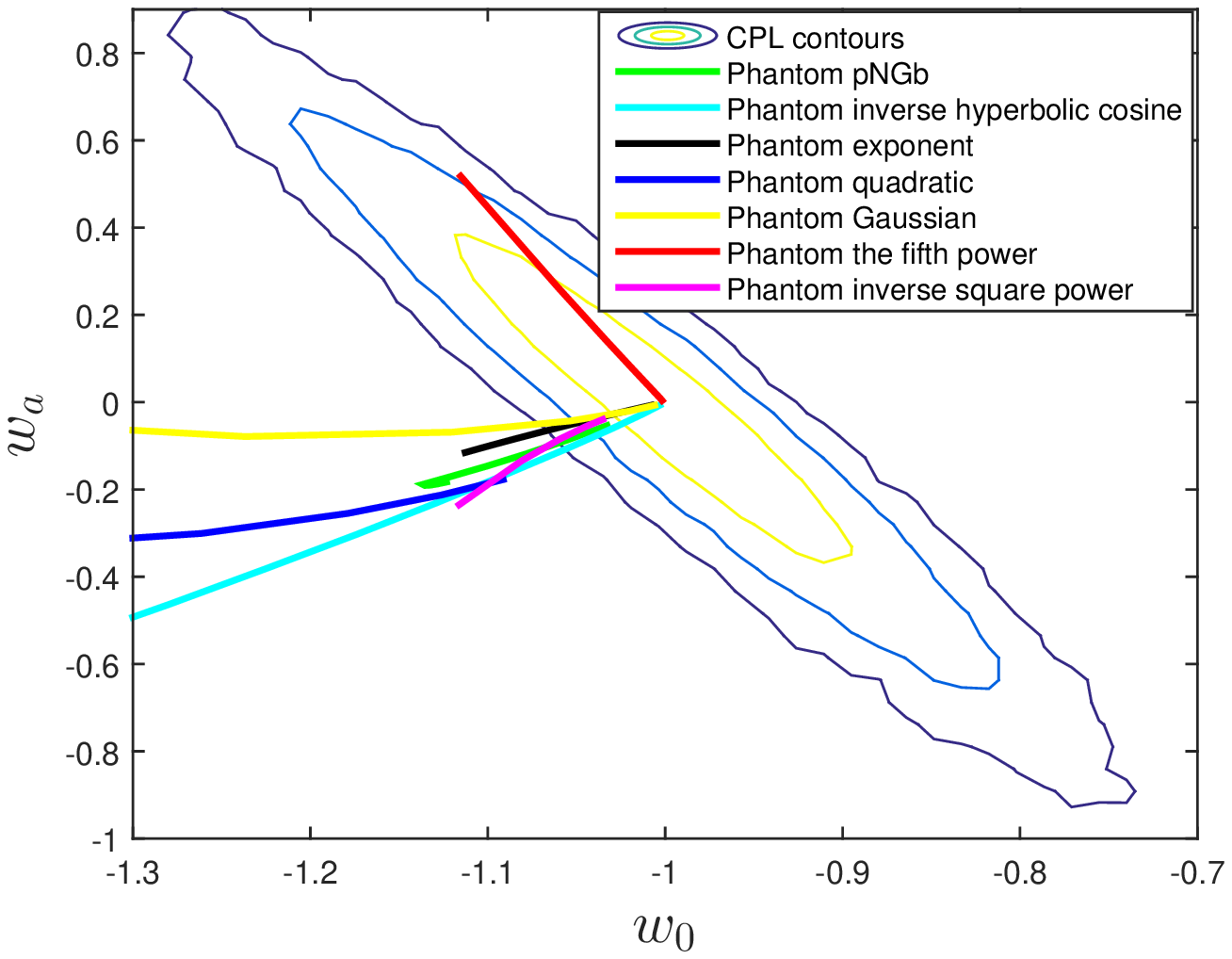}}
 \caption {Similar to the Fig.~(\ref{fig:f1}), for the phantom models.}
 \label{fig:f2}
 \end{figure}

As a additional exercise we looked at how well the CPL parametrization
approximates these dark energy models and where each dark energy model is
mapped in the CPL parameter space.  The CPL-$\Lambda$CDM contours in
Figs.~(\ref{fig:f1}-\ref{fig:f2}) represent 1, 2, and 3$\sigma$ confidence
levels for the CPL parametrization derived by fitting the same $H(z)$,
$d_A(z)$, and $f(a)\sigma_8$ data.  In order to check how well the CPL
parametrization Eq.~(\ref{eq:EOS}) describes the dark energy models, we find the
best-fit effective values of $w_0-w_a$ for a range of the free parameters of
each model. These results are presented in Fig.~(\ref{fig:f1}) for the
quintessence models and in Fig.~(\ref{fig:f2}) for the phantom models.  For an
easy visual representation of this information we pick a parameter with respect
to which the best fit $w_0$ and $w_a$ values are most sensitive and plot this
range within priors.

In Fig.(\ref{fig:f1}) we show that some of the dark energy models stay very
close to the $\Lambda$CDM for a wide range of parameter values within our
priors. The range of the EOS parameters for the Ferreira-Joyce, the inverse
exponent and the Sugra potentials is very small, it almost coincides with the
$\Lambda$CDM model EOS parameter ($w_0=-1, w_a=0$) consequently the likelihood
of these model parameters is relatively flat and they can only be distinguished
from $\Lambda$CDM model by Occam's razor type arguments. The Chang-Scherrer,
the Ur\~{e}na-L\'{o}pez-Matos, and the Barreiro potentials can result in up to
3$\sigma$ offsets from $\Lambda$CDM for some parameter values; the
Zlatev-Wang-Steinhardt, the Ratra-Peebles, the Albrecht-Skordis, and the Sahni-Wang potentials even extend beyond
3$\sigma$ confidence level.  This suggests that a significant fraction of the
parameter space can be distinguished based on posterior likelihood. All phantom
potentials in Fig.(\ref{fig:f2}), except the quadratic potential, exhibit a
similar behaviour. The quadratic potential lies outside the 3$\sigma$ contours
of projected DESI constraints. This happens because in this model it is
difficult to get a $\Lambda$CDM limit with a natural choice of parameter values
and initial conditions.

\section{Conclusions}
We have derived projected constraints on a number of dark energy models by
fitting them to a mock $H(z)$, $d_A(z)$, $f(a)\sigma_8(z)$ data generated in a
fiducial $\Lambda$CDM model. When fitting to predicted data one has to choose a
fiducial model. In our case this fiducial model was a Planck normalised
$\Lambda$CDM. While it is obvious that under this scheme the $\Lambda$CDM can
never be inferior to its alternatives, it is not clear a priori how strong the
evidence in favour of the $\Lambda$CDM model be. Our main goal was to see
whether the various Bayesian criteria would provide sufficient evidence in
favour of $\Lambda$CDM as opposed to considered alternatives. Our results seem
to suggest that even though all the scalar models have a $\Lambda$CDM limit
(which obviously remains a good fit to the data) DESI is capable of providing
enough evidence to reject them. Our conclusions come with the caveat that they
depend strongly on the adopted assumptions about the priors. The kinds of
Bayesian arguments that we employed can be very
sensitive to the assumed prior range of model parameters, Ref. \cite{Efstathiou:2008}. Our priors were reasonably wide and encompassed all of
the parameter space that was in the general $\Lambda$CDM ``neighbourhood'' and
thus compatible with currently available data. Physically motivated
restrictions on the parameter space would make the rejection of the alternative
models more difficult.

\begin{figure*}[h!]
\begin{center}
\resizebox{\textwidth}{!}{
\includegraphics{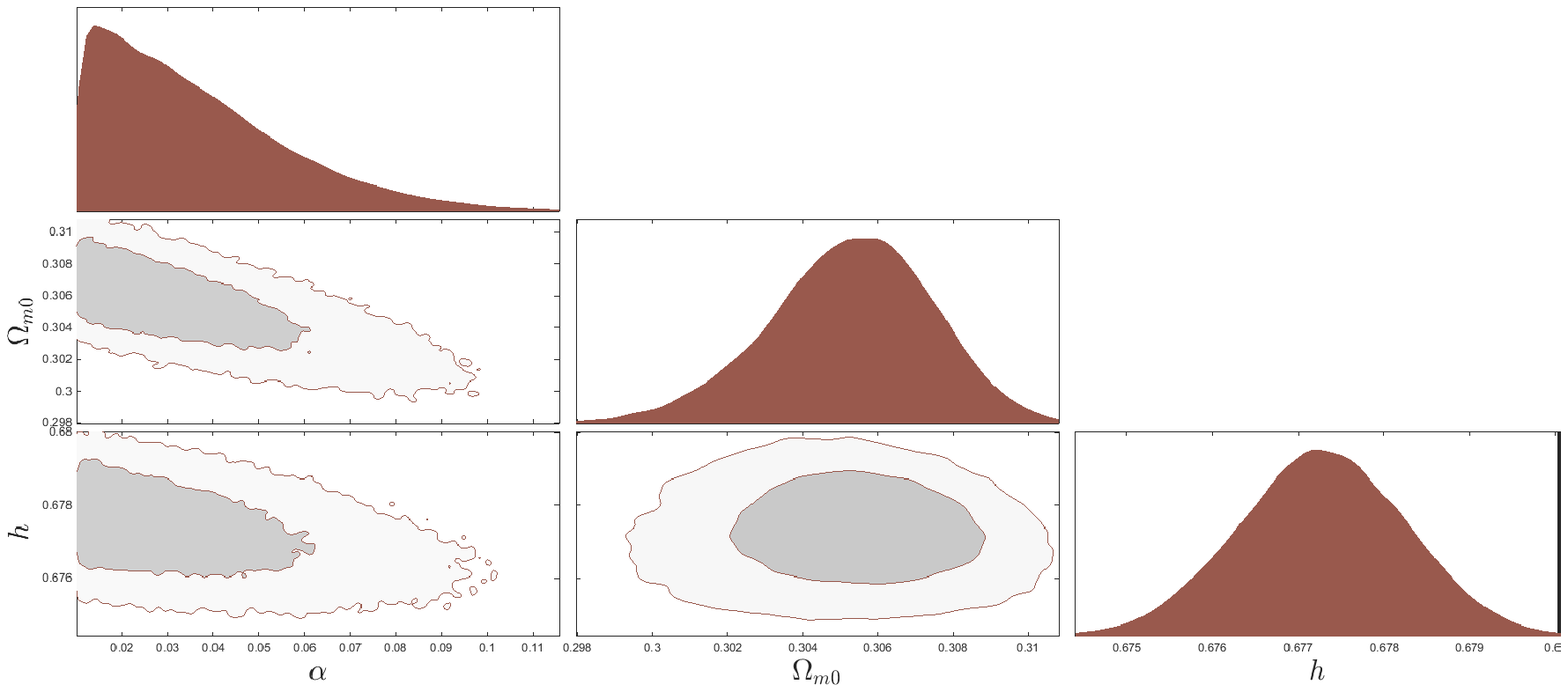}}
 \caption{The 2$\sigma$ confidence level contour plots for  various pairs of the free
parameters ($\alpha$, $\Omega_{\rm m0}$, $h$) for which the
$\phi$CDM model with the Ratra-Peebles potential
$V(\phi)=V_0M_\mathrm{pl}^2\phi^{-\alpha}$ is in the best fit with the $\Lambda$CDM model.}
 \label{fig:f3}
 \end{center}
\end{figure*}

\begin{figure*}[h!]
\begin{center}
\resizebox{\textwidth}{!}{
\includegraphics{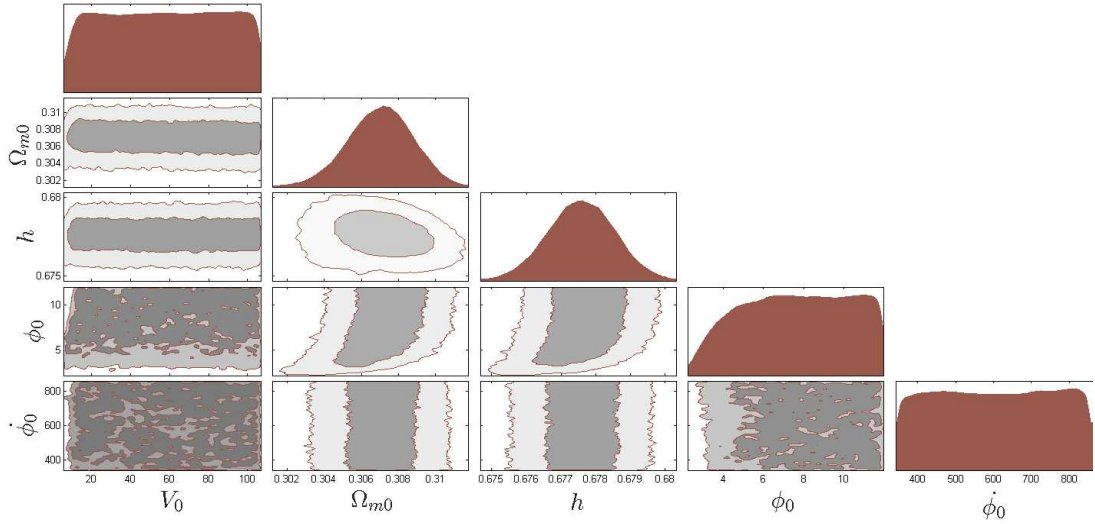}}
\caption{The 2$\sigma$ confidence level contour plots for  various pairs of the free
parameters ($V_0$, $\Omega_{\rm m0}$, $h$, $\phi_0$, $\dot{\phi_0}$) for which the  $\phi$CDM model with
the Zlatev-Wang-Steinhardt potential $V(\phi)=V_0(\exp({M_{\rm pl}/\phi})-1)$ is in the best
fit with the $\Lambda$CDM model.}
  \label{fig:f4}
   \end{center}
\end{figure*}

\begin{figure*}[h!]
\begin{center}
\resizebox{\textwidth}{!}{
\includegraphics{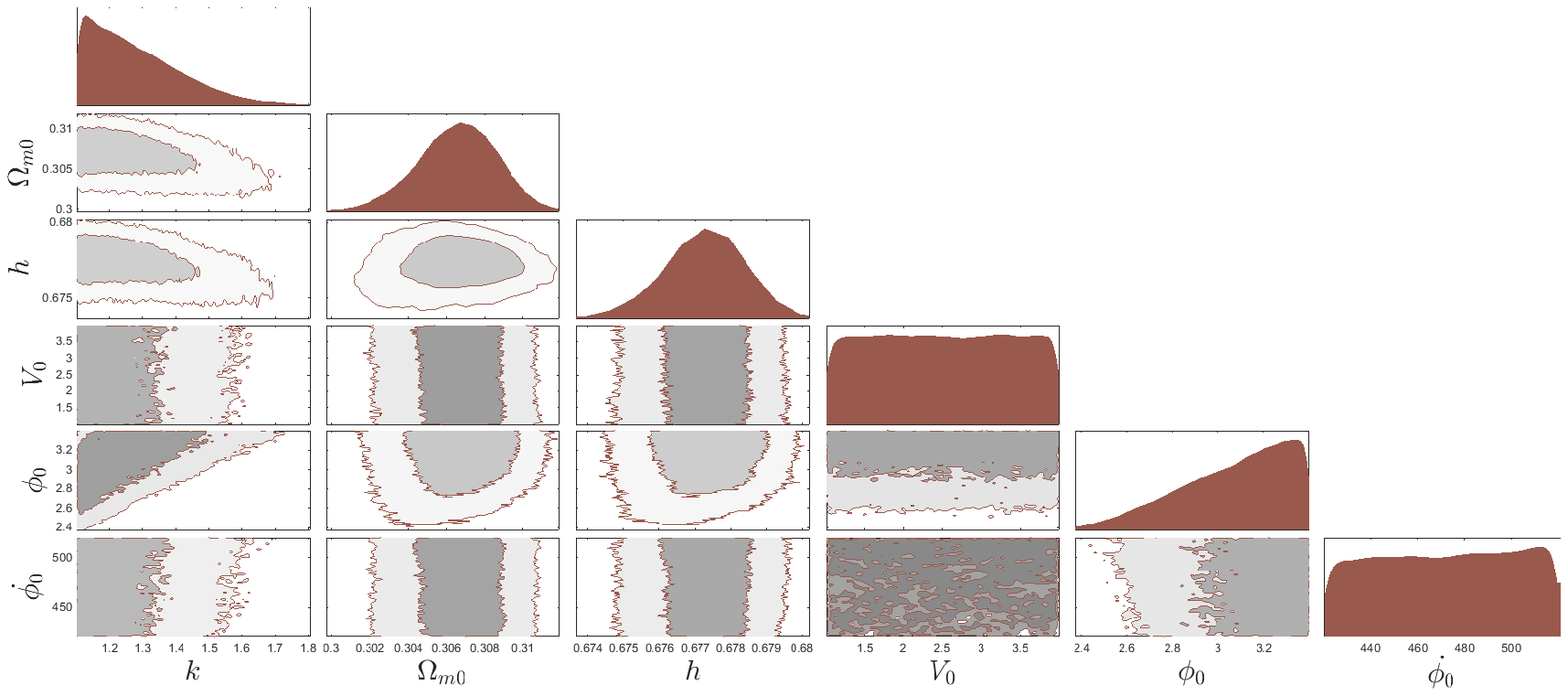}}
 \caption{The 2$\sigma$ confidence level contour plots for  various pairs of the free parameters ($k$, $\Omega_{\rm m0}$, $h$, $V_0$, $\phi_0$, $\dot{\phi_0}$) for which the $\phi$CDM model with the phantom Pseudo Nambu-Goldstone boson potential $V(\phi)=V_0(1-\cos(\phi/\kappa))$ is in the best fit with the $\Lambda$CDM model.}
 \label{fig:f5}
 \end{center}
\end{figure*}

In Figs.~(\ref{fig:f3}-\ref{fig:f5}) we show examples of the
constraints that we obtain for the quintessence Ratra-Peebles, the Zlatev-Wang-Steinhardt potentials
and for the phantom Pseudo Nambu-Goldstone potential. Since all models have
the $\Lambda$CDM model as their limit, strictly speaking it is impossible to
rule them out based on the likelihood arguments alone. Therefore we also used
commonly cited model comparison criteria in the Bayesian statistics such as the
Bayes factor, the $AIC$ and $BIC$ information criteria. Computing $AIC$ and
$BIC$ in our setup is straightforward. Since all models have the same maximum
likelihood by the construction the $AIC$ and the $BIC$ become simply functions
of the number of the extra parameters. To compute the Bayes factors we
integrated the posterior within the bounds given in the Tables \ref{table:QP}
and \ref{table:PP}. The results of the $AIC$, $BIC$, and Bayes factors for all
the dark energy models are summarized in the  Tables \ref{table:bicq} and
\ref{table:bicp}. These numbers clearly demonstrated that if the $\Lambda$CDM
model is the true description of dark energy, the full DESI data will be able
to strongly discriminate most scalar field dark energy models currently under
consideration.  These results however need to be taken with a grain of salt.
The evidence values are very sensitive to the prior ranges. We only restricted
the prior range based on constraints, by using the phenomenological method
developed by us. Further restriction of the parameter ranges could
significantly increase the evidence value. The results were derived assuming a
fiducial $\Lambda$CDM model and the low value of evidence simply means that the
model would be easier to discriminate if $\Lambda$CDM was the true model. The
flip side of this is that if instead the dynamic dark energy models were true
that would also show up more obviously in the data.

We also explored how the dark energy models are map-\\ped to the CPL
parameter surface. For the models considered in our work this parametrization
seems to work reasonably well even for the wide redshift range in a sense that
the model predictions are always within one percent of of corresponding CPL
predictions.

\begin{acknowledgement}
{ \bf Acknowledgements}
It is our great pleasure to thank Bharat Ratra for useful comments and
discussions.  We appreciate discussions with Natalia Arhipova, Leonardo
Campanelli, Vasil Kukhianidze, Olga Navros, Bohdan Novosyadlyj, and Alexander
Tevzadze. Support through the Shota Rustaveli Georgian NSF (grants FR/339/6-350/14 and PhD\_F\_17\_196), the CRDF-SRNSF-GRDF Georgia Women's Research Fellowship Program grant WRF-14-22, the Swiss NSF SCOPES (grant IZ 7370-152581), and the NSF Astrophysics and Astronomy Grant (AAG) Program
(grant AST-1615940), DOE grant DEFG 03-99EP41093, and NASA grant
12-EUCLID11-0004 are gratefully acknowledged. TK also thanks the High Energy
and Cosmology division and the Associate Membership Program at International
Center for Theoretical Physics (ICTP) for hospitality and partial support.
OA thanks to KSU Physics department for hospitality.
\end{acknowledgement}


\begin{thebibliography}{99}

\bibitem{Riess:1998cb}
  A.~G.~Riess {\it et al.} [Supernova Search Team],
  Astron.\ J.\  {\bf 116}, 1009 (1998).

\bibitem{Perlmutter:1998np}
  S.~Perlmutter {\it et al.} [Supernova Cosmology Project Collaboration],
  Astrophys.\ J.\  {\bf 517}, 565 (1999).

\bibitem{Komatsu:2010fb}
  E.~Komatsu {\it et al.} [WMAP Collaboration],
  Astrophys.\ J.\ Suppl.\  {\bf 192}, 18 (2011).

\bibitem{Frieman:2008sn}
  J.~Frieman, M.~Turner and D.~Huterer,
  Ann.\ Rev.\ Astron.\ Astrophys.\  {\bf 46}, 385 (2008).

\bibitem{Weinberg:2012es}
  D.~H.~Weinberg, M.~J.~Mortonson, D.~J.~Eisenstein, C.~Hirata, A.~G.~Riess and E.~Rozo,
  Phys.\ Rept.\  {\bf 530}, 87 (2013).

\bibitem{Sahni:1999gb}
  V.~Sahni and A.~A.~Starobinsky,
  Int.\ J.\ Mod.\ Phys.\ D {\bf 9}, 373 (2000).

\bibitem{Carroll:2000fy}
  S.~M.~Carroll,
  Living Rev.\ Rel.\  {\bf 4}, 1 (2001).

\bibitem{Sahni:2002kh}
  V.~Sahni,
  Class.\ Quant.\ Grav.\  {\bf 19}, 3435 (2002).

\bibitem{Padmanabhan:2002ji}
  T.~Padmanabhan,
  Phys.\ Rept.\  {\bf 380}, 235 (2003).

\bibitem{Peebles:2002gy}
  P.~J.~E.~Peebles and B.~Ratra,
  Rev.\ Mod.\ Phys.\  {\bf 75}, 559 (2003).

\bibitem{Martin:2012bt}
  J.~Martin,
  Comptes Rendus Physique {\bf 13}, 566 (2012).

\bibitem{Padilla:2015aaa}
  A.~Padilla,
  arXiv:1502.05296 (2015).

\bibitem{Copeland:2006wr}
  E.~J.~Copeland, M.~Sami and S.~Tsujikawa,
  Int.\ J.\ Mod.\ Phys.\ D {\bf 15}, 1753 (2006).

\bibitem{Caldwell:2009ix}
  R.~R.~Caldwell and M.~Kamionkowski,
  Ann.\ Rev.\ Nucl.\ Part.\ Sci.\  {\bf 59}, 397 (2009).

\bibitem{Cai:2009zp}
  Y.~F.~Cai, E.~N.~Saridakis, M.~R.~Setare and J.~Q.~Xia,
  Phys.\ Rept.\  {\bf 493}, 1 (2010)

\bibitem{Bahamonde:2017ize}
  S.~Bahamonde, C.~G.~Boehmer, S.~Carloni, E.~J.~Copeland, W.~Fang and N.~Tamanini,
  arXiv:1712.03107 (2017)

  \bibitem{Font-Ribera:2013rwa}
  A.~Font-Ribera, P.~McDonald, N.~Mostek, B.~A.~Reid, H.~J.~Seo and A.~Slosar,
  JCAP {\bf 1405}, 023 (2014)

\bibitem{Levi:2013gra}
  M.~Levi {\it et al.} [DESI Collaboration],
  arXiv:1308.0847 (2013)

\bibitem{Aghamousa:2016zmz}
  A.~Aghamousa {\it et al.} [DESI Collaboration],
  arXiv:1611.00036 (2016)

  \bibitem{Amendola:2012ys}
  L.~Amendola {\it et al.} [Euclid Theory Working Group],
  Living Rev.\ Rel.\  {\bf 16}, 6 (2013)

 \bibitem{Spergel:2015sza}
  D.~Spergel {\it et al.},
  arXiv:1503.03757 (2015)

\bibitem{Starobinsky:1978}
 A.~A.~Starobinsky,
 Sov.\ Astron.\ Lett. {\bf 4}, 82 (1978)

\bibitem{Ratra:1987rm}
  B.~Ratra and P.~J.~E.~Peebles,
  Phys.\ Rev.\ D {\bf 37}, 3406 (1988)

\bibitem{Steinhardt:1999nw}
  P.~J.~Steinhardt, L.~M.~Wang and I.~Zlatev,
  Phys.\ Rev.\ D {\bf 59}, 123504 (1999)

\bibitem{Linder:2007wa}
  E.~V.~Linder,
  Gen.\ Rel.\ Grav.\  {\bf 40}, 329 (2008)

\bibitem{Ludwick:2017tox}
  K.~J.~Ludwick,
  Mod.\ Phys.\ Lett.\ A {\bf 32}, 28, 1730025 (2017)

\bibitem{Ferreira:1997hj}
  P.~G.~Ferreira and M.~Joyce,
  Phys.\ Rev.\ D {\bf 58}, 023503 (1998)

 \bibitem{Zlatev:1998tr}
  I.~Zlatev, L.~M.~Wang and P.~J.~Steinhardt,
  Phys.\ Rev.\ Lett.\  {\bf 82}, 896 (1999)

\bibitem{Brax:1999gp}
  P.~Brax and J.~Martin,
  Phys.\ Lett.\ B {\bf 468}, 40 (1999)

\bibitem{Sahni:1999qe}
  V.~Sahni and L.~M.~Wang,
  Phys.\ Rev.\ D {\bf 62}, 103517 (2000)

\bibitem{Barreiro:1999zs}
  T.~Barreiro, E.~J.~Copeland and N.~J.~Nunes,
  Phys.\ Rev.\ D {\bf 61}, 127301 (2000)

\bibitem{Albrecht:1999rm}
  A.~Albrecht and C.~Skordis,
  Phys.\ Rev.\ Lett.\  {\bf 84}, 2076 (2000)

\bibitem{UrenaLopez:2000aj}
  L.~A.~Urena-Lopez and T.~Matos,
  Phys.\ Rev.\ D {\bf 62}, 081302 (2000)

\bibitem{Caldwell:2005tm}
  R.~R.~Caldwell and E.~V.~Linder,
  Phys.\ Rev.\ Lett.\  {\bf 95}, 141301 (2005)

\bibitem{Chang:2016aex}
  H.~Y.~Chang and R.~J.~Scherrer,
  arXiv:1608.03291 (2016) 

\bibitem{Scherrer:2008be}
  R.~J.~Scherrer and A.~A.~Sen,
  Phys.\ Rev.\ D {\bf 78}, 067303 (2008)

\bibitem{Dutta:2009dr}
  S.~Dutta and R.~J.~Scherrer,
  Phys.\ Lett.\ B {\bf 676}, 12 (2009)

\bibitem{Frieman:1995pm}
  J.~A.~Frieman, C.~T.~Hill, A.~Stebbins and I.~Waga,
  Phys.\ Rev.\ Lett.\  {\bf 75}, 2077 (1995)

\bibitem{Rakhi:2009qf}
  R.~Rakhi and K.~Indulekha,
  arXiv:0910.5406 (2009)

\bibitem{Bag:2017vjp}
  S.~Bag, S.~S.~Mishra and V.~Sahni,
  JCAP {\bf 1808}, 8, 009 (2018)

\bibitem{Suzuki:2011hu}
  N.~Suzuki {\it et al.},
  Astrophys.\ J.\  {\bf 746}, 85 (2012)

\bibitem{Betoule:2014frx}
  M.~Betoule {\it et al.} [SDSS Collaboration],
  Astron.\ Astrophys.\  {\bf 568}, A22 (2014)

\bibitem{Ade:2013zuv}
  P.~A.~R.~Ade {\it et al.} [Planck Collaboration],
  Astron.\ Astrophys.\  {\bf 571}, A16 (2014)

\bibitem{Ade:2015xua}
  P.~A.~R.~Ade {\it et al.} [Planck Collaboration],
  Astron.\ Astrophys.\  {\bf 594}, A13 (2016)

\bibitem{Novosyadlyj:2012qu}
  B.~Novosyadlyj, O.~Sergijenko, R.~Durrer and V.~Pelykh,
  JCAP {\bf 1306}, 042 (2013)

\bibitem{Heavens:2017hkr}
  A.~Heavens, Y.~Fantaye, E.~Sellentin, H.~Eggers, Z.~Hosenie, S.~Kroon and A.~Mootoovaloo,
  Phys.\ Rev.\ Lett.\  {\bf 119}, 10, 101301 (2017)

\bibitem{Dhawan:2017leu}
  S.~Dhawan, A.~Goobar, E.~Mörtsell, R.~Amanullah and U.~Feindt,
  JCAP {\bf 1707}, 07, 040 (2017)

\bibitem{Lonappan:2017lzt}
  A.~I.~Lonappan, S.~Kumar, Ruchika, B.~R.~Dinda and A.~A.~Sen,
  Phys.\ Rev.\ D {\bf 97}, 4, 043524 (2018)

\bibitem{Weinberg:2008zzc}
  S.~Weinberg, ``Cosmology,''
  Oxford, UK: Oxford Univ. Pr.  593  (2008)

\bibitem{Peebles:1987ek}
  P.~J.~E.~Peebles and B.~Ratra,
  Astrophys.\ J.\  {\bf 325}, L17 (1988)

\bibitem{Yoo:2012ug}
  J.~Yoo and Y.~Watanabe,
  Int.\ J.\ Mod.\ Phys.\ D {\bf 21}, 1230002 (2012)

\bibitem{LM1985}
  F.~Lucchin and S.~Matarrese,  R.~J.~Scherrer and A.~A.~Sen,
  Phys.\ Rev.\ D {\bf 32}, 1316 (1985)

\bibitem{Pace:2010sn}
  F.~Pace, J.~-C.~Waizmann and M.~Bartelmann,
  Mon. Not. Roy. Astron. Soc. {\bf 406}, 1865 (2010)

\bibitem{LHuillier:2017ani}
  B.~L'Huillier, A.~Shafieloo and H.~Kim,
  Mon.\ Not.\ Roy.\ Astron.\ Soc.\  {\bf 476}, 3, 3263 (2018)

\bibitem{Chevallier:2000qy}
  M.~Chevallier and D.~Polarski,
  Int.\ J.\ Mod.\ Phys.\ D {\bf 10}, 213 (2001)

\bibitem{Linder:2002et}
  E.~V.~Linder,
  Phys.\ Rev.\ Lett.\  {\bf 90}, 091301 (2003)

\bibitem{Barger:2006vc}
  V.~Barger, Y.~Gao and D.~Marfatia,
  Phys.\ Lett.\ B {\bf 648}, 127 (2007).

\bibitem{Wang:1998gt}
  L.~M.~Wang and P.~J.~Steinhardt,
  Astrophys.\ J.\  {\bf 508}, 483 (1998)

\bibitem{Polarski:2016ieb}
  D.~Polarski, A.~A.~Starobinsky and H.~Giacomini,
  JCAP {\bf 1612}, no. 12, 037 (2016)

\bibitem{Avsajanishvili:2015elt}
  O.~Avsajanishvili, L.~Samushia, N.~A.~Arkhipova and T.~Kahniashvili,
  Astron. \& Astrophys. Transactions, {\bf 30}, 01, 95 (2017)


\bibitem{Wu:2009zy}
  P.~Wu, H.~W.~Yu and X.~Fu,
  JCAP {\bf 0906}, 019 (2009)

\bibitem{Linder:2005in}
  E.~V.~Linder,
  Phys.\ Rev.\ D {\bf 72}, 043529 (2005)

\bibitem{Linder:2007hg}
  E.~V.~Linder and R.~N.~Cahn,
  Astropart.\ Phys.\  {\bf 28}, 481 (2007)

  \bibitem{Arkhipova:2014dha}
  O.~Avsajanishvili, N.~A.~Arkhipova, L.~Samushia and T.~Kahniashvili,
  Eur.\ Phys.\ J.\ C {\bf 74}, 11, 3127 (2014)

\bibitem{Sangwan:2017kxi}
  A.~Sangwan, A.~Mukherjee and H.~K.~Jassal,
  JCAP {\bf 1801}, 1, 018 (2018)

\bibitem{Akaike:1974}
  Hirotugu~Akaike,
 IEEE Transactions on Automatic Control  {\bf 19}, 06, 716 (1974)

\bibitem{Schwarz:1978}
 Gideon~ E.~Schwarz,
 Annals of Statistics  {\bf 06}, 461 (1978)

\bibitem{Efstathiou:2008}
 G.~Efstathiou,
 Monthly Notices of the Royual Astronomical Society, 388, 1314 (2008)


\end{thebibliography}
\end{document}